\documentclass[a4paper]{jpconf}
\usepackage{graphicx}
\usepackage{bbold}
\newcommand{\be}{\begin{eqnarray}}
\newcommand{\ee}{\end{eqnarray}}

\newcommand{\ox}{\overline{x}}

\newcommand{\veff}{V_{eff}}    

\begin{document}
\title{Investigating Biological Matter with Theoretical Nuclear Physics Methods}

\author{Pietro Faccioli}
\address{Dipartimento di Fisica, Universit\`a degli Studi di Trento, Via Sommarive 14, Povo (Trento), I-38123 Italy.}
\address{INFN, Gruppo Collegato di Trento, Via Sommarive 14, Povo (Trento), I-38123 Italy.} 

\ead{faccioli@science.unitn.it}

\begin{abstract}
The internal dynamics of strongly interacting systems  and that of bio-polymers such as proteins display several important analogies, despite the huge difference in their characteristic 
energy and length scales. 
For example, in all such systems, collective excitations, cooperative transitions and phase transitions emerge as the result of the interplay of strong correlations with quantum or thermal 
fluctuations. 
In view of such an observation, some theoretical methods initially developed in the context of theoretical nuclear physics have been adapted to investigate the dynamics of biomolecules.   
In this talk, we review some of our recent studies performed along this direction. In particular, we discuss how the path integral formulation of the molecular dynamics allows to overcome some 
 of the long-standing problems and limitations which emerge when simulating the protein folding dynamics at the atomistic level of detail. 
  \end{abstract}

\section{Introduction}

\begin{figure}[t]
\center{
  \includegraphics[width=12 cm]{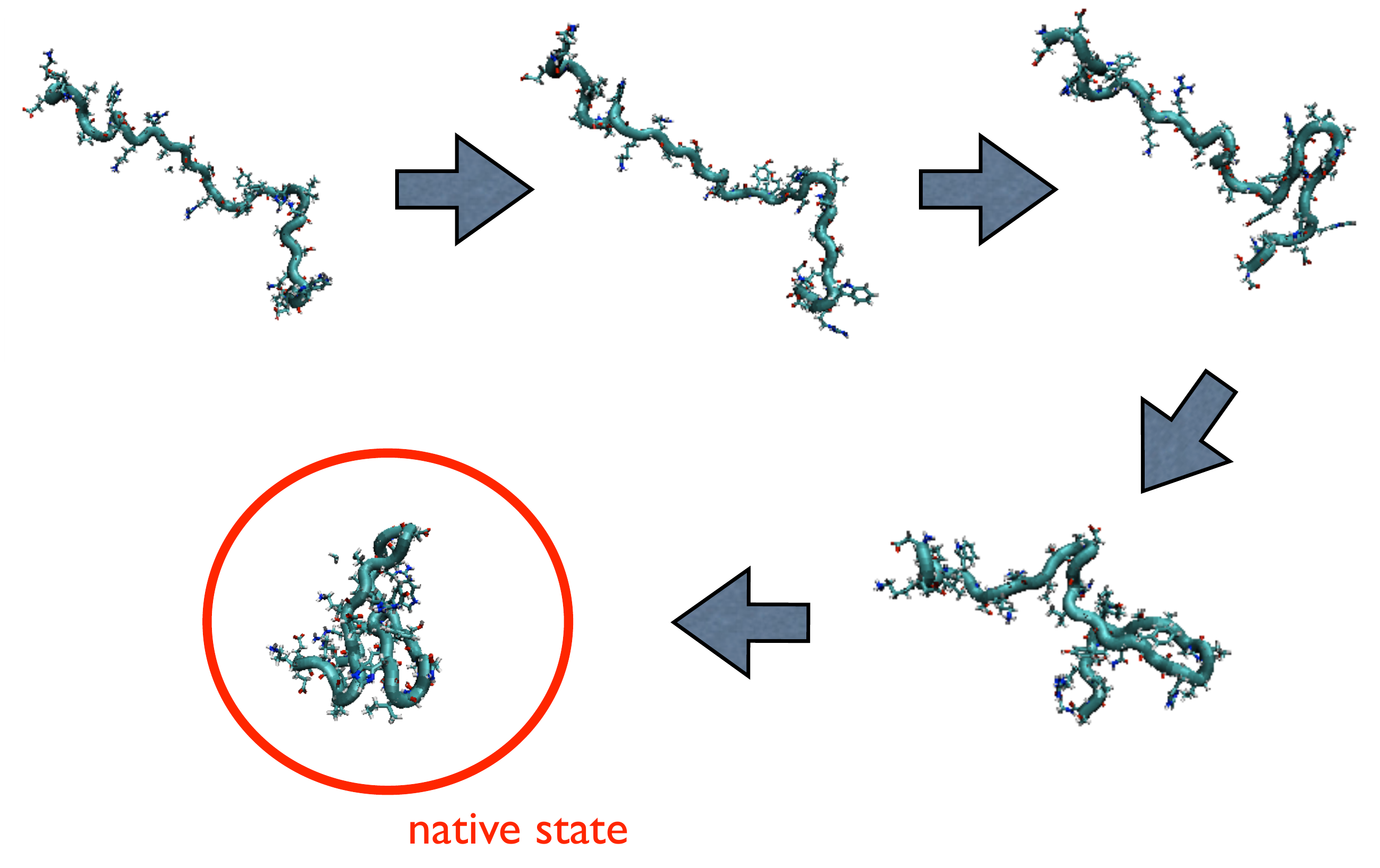}
  \caption{Schematic representation of a protein folding pathway. This trajectory corresponds to a biological instanton and was obtained using the
   Dominant Reaction Pathways approach (see discussion in section \ref{DRP}. }}
  \label{FoldingFIP}
  \end{figure}

Proteins are  macromolecules involved in virtually all the biochemical processes which take place inside the cell, such as e.g. catalysis, charge transport, 
signal transduction. Clearly, understanding the physical principles which drive the internal dynamics of individual proteins and shape the protein-protein interaction would have countless implications in molecular biology, drug design and nano-biotechnology. In particular, a central open challenge at the interface of physics, biochemistry and molecular
 biology concerns the prediction of the protein folding pathways i.e. the sequence of conformational
 transitions which these molecules perform in order to reach their stable and biologically active configuration (see e.g. Fig. \ref{FoldingFIP}) . 

From a theoretical perspective, the problem of determining the time evolution of proteins has been mostly investigated by means of Molecular Dynamics (MD) simulations, i.e. through the 
direct integration of the classical equations of motion of all the atoms in the system. From the conceptual point of view, such an approach should
 provide a reasonable approximation, since quantum corrections are expected to be 
quite small.  On the other hand, from the practical point of view, the applicability of classical MD simulations is strongly limited by their intrinsic computational complexity. Indeed,
 even using computational resources comparable with those employed in the largest contemporary lattice QCD simulations, it is possible to follow  
the dynamics over time intervals which do not exceed a few hundreds ns~\cite{Freddolino}.  Unfortunately, the dynamics  which drives the protein phenomenology  occurs at time scales which are 
several orders of
 magnitude larger, typically ranging from few ms to several seconds. 
 
 The decoupling of the microscopic time scales from the time scales at which conformational reactions occur is a common feature in many bio-molecular systems. It is due to the fact that most 
 conformational
 transitions are cooperative and involve the overcoming of free-energy barriers. As a result, in a MD simulation,  most computational time is wasted to 
 describe thermal oscillations inside the initial meta-stable state.
 
Given the huge gap between the time scales at which the relevant physics takes place and the time intervals which can be covered by 
 MD simulations, it is natural to expect that a real breakthrough in our comprehension of biological matter at the fundamental level is more likely to come from new ideas and new 
 theoretical approaches, rather than from the next several generations of supercomputers. 

In the next sections,  we discuss how functional integral approaches originally developed in the context of nuclear theory can provide an extremely  powerful framework
 to investigate the dynamics of proteins as well as other complex molecular systems. Before introducing the mathematical details
 of this theory, it is instructive to discuss some basic aspects of protein structure and phenomenology and to highlight several analogies with
the dynamics of strongly interacting systems.
  
\section{QCD and Protein Dynamics: So Far Yet So Close}

 From the biochemical  point of view, proteins are poly-peptide chains, i.e. weakly branched polymers formed by 20 different types of monomers called amino-acids.
   Typical globular proteins consist of about 100 amino-acids, but some chains can contain as many as $\sim 30,000$ of such monomers.
 
\begin{figure}[t]
\center{
  \includegraphics[width=9 cm]{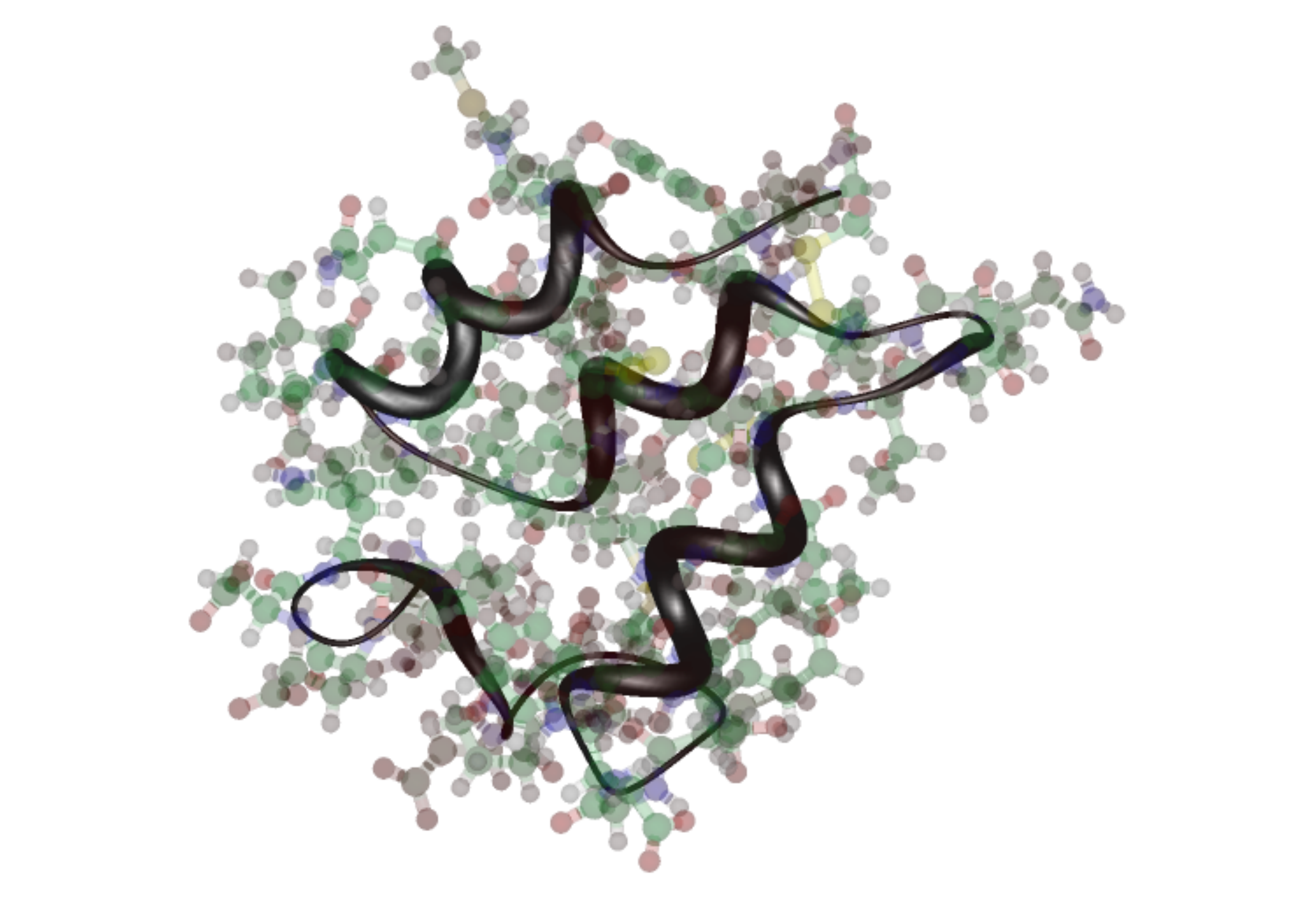}
  \caption{Ball-and-stick representation of the native state of a globular protein. The tube represents the back-bone of the chain. }}
  \label{native}
  \end{figure}

 From the physical point of view~\cite{proteinphysics}, proteins are very different from essentially all other known polymers: indeed, while standard  hetero-polymers 
 are prototypes of disordered mesoscale systems, proteins
are characterized by their unique "ability" to spontaneously and reversibly fold into a well-defined 
stable configuration (denominated the \emph{native state}), in which they perform their biological functions (see e.g. Fig. \ref{native}). 
Such a remarkable feature is the result of the evolutionary pressure, which has selected only a very specific subset in the space of all possible  amino-acid sequences. 

\begin{figure}[t]
\center{
  \includegraphics[width=12 cm]{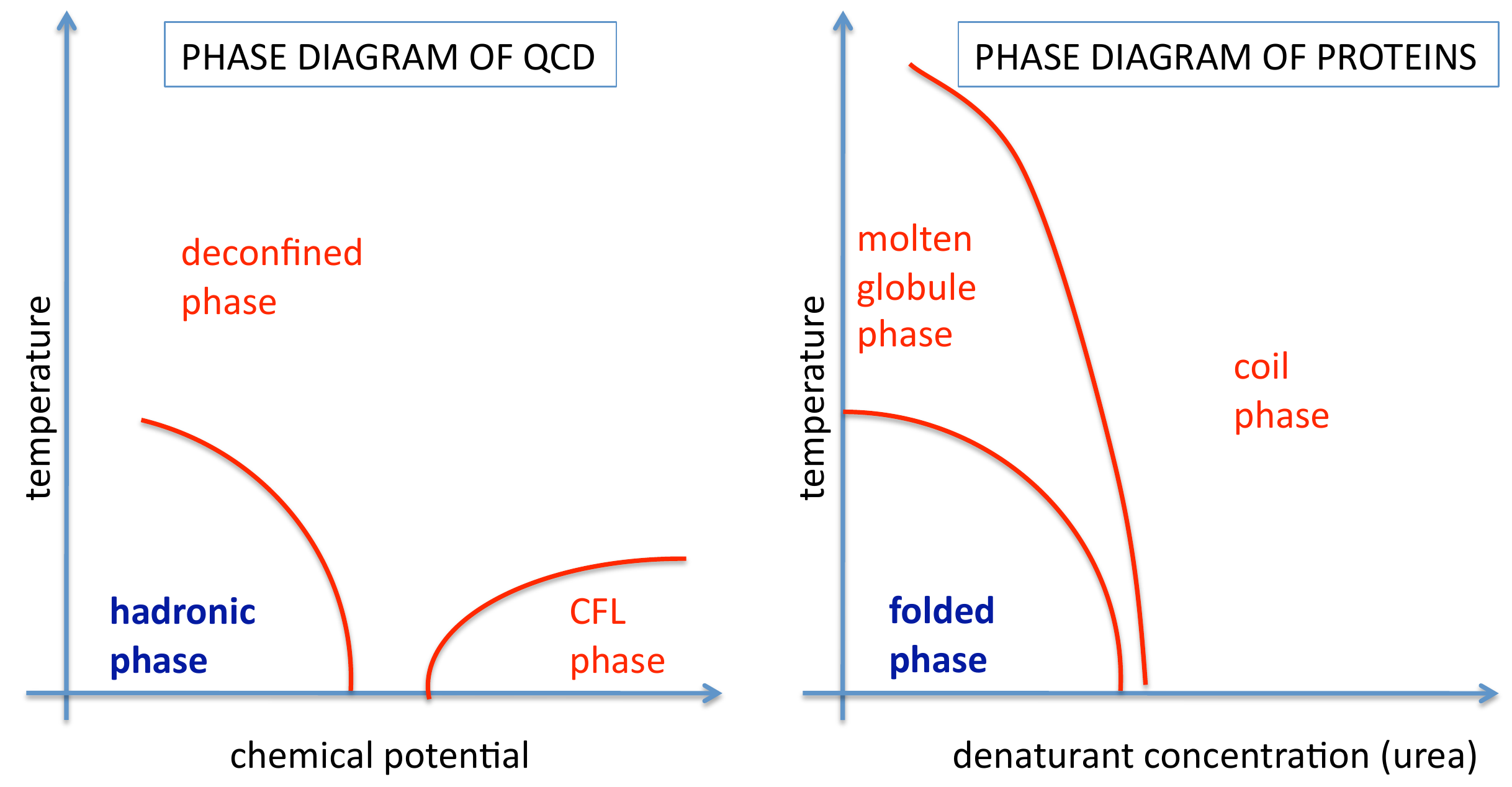}
  \caption{Comparison between the phase diagram of QCD and of proteins in solution. }}
  \label{phase}
  \end{figure}

The native state of a protein is characterized by an extremely low entropy, since it is defined by the small thermal oscillations around a single conformation. 
In this sense,  proteins may be regarded as 
non-symmetric mesoscale crystals.  If the temperature of the water 
surrounding the molecule is increased, or if a denaturant agent 
(typically urea) is added to the solvent, then proteins can reach a much more entropic phase called the \emph{molten globule}. In this state, the chain is still collapsed, but is free to explore the different 
compact configurations, in analogy with a liquid drop. Finally, at even 
higher temperatures or denaturant concentrations, proteins swell and start to behave as flexible random coils (see right panel of Fig. \ref{phase}).

Interestingly, the phase diagram of proteins in solution displays several analogies with that of strongly interacting matter (see  left panel of Fig. \ref{phase}). 
Indeed, the  low-temperature and 
low-density phase of  QCD is characterized by a very low entropy, since only colorless hadrons are present in the spectrum. On the other hand, above the
 deconfinement phase transition,  the entropy is much larger, since also quantum states in color multiplets can contribute to the partition function. 
 
In addition, flavor symmetry implies that the hadronic spectrum can be approximatively organized into SU(3) multiplets. Hence, since quantum states belonging to the same multiplet can be transformed 
into one another by a flavor rotation, the number of truly distinct hadronic wave-functions is actually much smaller than the number of hadrons recorded in the particle data book.
Similarly, it has been experimentally observed that nearly identical native structures are adopted by poly-peptide chains characterized by very different amino-acid sequences.
This is to say that the number of native structures which can be realized is much smaller than the number of sequences in the data bank of known proteins. 
   
These analogies are not accidental. They reflect the fact that the dynamics of both hadrons and proteins is characterized by the interplay of large
correlations and quantum or thermal fluctuations. It is natural to address the question whether some of the powerful theoretical techniques developed 
to describe the non-perturbative internal dynamics of hadrons  may be exported to describe
the  non-perturbative internal dynamics of  proteins. In the next section, we shall see that this intuition can be made rigorous, i.e. that it is possible to formulate the dynamics of
 biomolecular systems in terms of an effective quantum many-body problem. 

\section{Path Integral Formulation of Molecular Dynamics}

The dynamics of biomolecular systems in solution is often described by means of the Langevin equation, which reads
\begin{equation}
m~\ddot{x}=- \gamma \dot{x} - \nabla U(x) + \xi(t),
\label{Lan1}
\end{equation}
In this equation,  $x=(x_1,x_2,...,x_N)$ is a macro-vector specifying the coordinates of all the atomic nuclei in the molecule, $U(x)$ is the potential energy 
and $\xi(t)$ is a stochastic force, which simulates the effects of the random Brownian collisions with the  molecules in the solvent.
The contemporary models for the potential energy $U(x)$ --- which is often improperly called the \emph{force field}~--- are defined  phenomenologically  in order to 
reproduce the results of quantum chemistry calculations of the equilibrium configurations of small molecules \cite{Leach}.  
For sake of simplicity and without loss of generality, in the following we present the formalism in the case of one particle in one spacial dimension. The generalization to the multi-dimensional case is straightforward.
  
The acceleration term on the left hand side of  Eq. (\ref{Lan1}) introduces inertial effects which are damped on a time-scale $t\sim m/\gamma\equiv \tau_D$.  
For most biomolecular systems,  $\tau_D$ is of  order of the fraction of a $ps$, hence much smaller than the relevant microscopic time-scale associated to the 
 dynamics of the torsional angles, which takes place at the $ns$ time-scale.

For $t\gg\tau_D$ the Langevin Eq. (\ref{Lan1}) reduces to
\begin{equation}
\frac{\partial x}{\partial t} = -\frac{D}{k_B T} ~ \nabla U(x) + \eta(t),
\label{Lan2}
\end{equation}
where  $D=\frac{k_B T}{ m \gamma}$ is the so-called diffusion coefficient and $\eta(t)= \frac{1}{\gamma}~\xi(t)$ is a rescaled white noise that satisfy the so-called
 fluctuation-dissipation relationship:
\be
\langle\eta(t)\eta(t')\rangle= 2 D \delta(t-t').
\ee

It can be shown ---see e.g. \cite{Schwabl}--- that the probability distribution sampled by the stochastic differential Eq. (\ref{Lan2}) satisfies then the 
Fokker-Planck (FP) equation:
\be
\frac{  \partial}{  \partial\,t}~P(x,t)= D \nabla 
\left(\frac{1}{k_B T}~\nabla\,U(x)~P(x,t)~\right)
+ D\,\nabla^2~P(x,t).
\label{FPE}
\ee
A universal property of all the solutions of the Eq. (\ref{FPE}) is that, in the long time limit, they converge to the Boltzmann weight, regardless of their initial conditions.

By performing the substitution 
\be
P(x,t) \equiv e^{-\frac{1}{2 k_B T} U(x)}~\psi(x,t),
\ee
the FP Eq. (\ref{FPE}) can be formally re-written as a Schr\"odinger Equation in imaginary time:
\be
 -\frac{\partial}{\partial t} \psi(x,t) = \hat{H}_{ef\!f}~\psi(x,t),
\label{SE}
\ee
where the effective "Quantum Hamiltonian" operator reads
\be
\hat{H}_{ef\!f}~=~- D \nabla^2 + V_{ef\!f}(x),
\label{Heff}
\ee
and $V_{eff}(x)$ is an effective potential and is defined as 
\be
V_{ef\!f}(x)= \frac{D}{4 (k_B T)^2}~ \left[ \left(\nabla  U(x)\right)^2 - 2 k_B T\, \nabla^2 U(x)\right].
\label{Veff}
\ee
Hence, we have shown that the problem of studying the \emph{real time} stochastic dynamics of a protein in solution can be mapped into the 
problem of determining the \emph{imaginary time} evolution of an effective quantum many-body system.

Based on the analogy with quantum mechanics, it is immediate to obtain a path-integral representation of the solution of (\ref{FPE}), subject to the boundary conditions 
$x(0)=x_i$ and $x(t)=x_f$:
\be
\label{PI}
P\left(x_f,t \left|x_i\right.\right) &=& e^{-\frac{1}{2 k_B T}(U(x_f)-U(x_i))} 
\langle x_i | e^{- \hat{H}_{ef\!f} t}| x_f\rangle  \nonumber \\
&=& e^{-\frac{1}{2 k_B T}(U(x_f)-U(x_i))}~\int_{x_i}^{x_f} 
\mathcal{D}x(\tau)~e^{- \int_{0}^{t} d\,\tau~ \left(\frac{\dot{x}^2(\tau)}{4 D}+ V_{ef\!f}[x(\tau)]\right)}.
\label{path2}
\ee
This equation expresses the fact that the Green's function of the  Fokker-Planck equation is formally equivalent to a quantum-mechanical propagator in imaginary time.
On the other hand, it is important to stress that the variable $t$ which enters in these equations corresponds to the physical time. 

\section{The Theory of Biological Instantons}
\label{DRP}
The path integral formulation of the stochastic dynamics of protein is particularly convenient if one is interested in investigating the structure of the protein folding pathways. 
Indeed, such a formalism allows to focus directly on the non-equilibrium dynamical trajectories which actually reach the native state $x_f$ in a given time interval $t$.  
This way, one  avoids wasting time to simulate trajectories which remain confined in the initial denatured state. A second important advantage of  Eq. (\ref{PI}) is that
it assigns a probability density
to each folding pathway~$x(\tau)$:
\be
\textrm{Prob.}[x(\tau)] \propto \exp[- S_{eff}[x]].
\ee 
 
 Clearly, the folding pathways which are most likely 
to be realized by a given protein are the solutions of the "classical" equations of motion generated by the effective action
\be\label{Seff}
 S_{eff}[x]= \int_{0}^t d\tau ~\left(~\frac{\dot{x}^2}{4 D} + V_{eff}[x]~\right).
 \ee
 These trajectories typically  overcome barriers of the effective potential $V_{eff}(x)$. Hence, they correspond to instantons in the effective quantum theory.  The so-called Dominant Reaction Pathways (DRP)
approach developed by our group \cite{DRPtheory1, DRPtheory1.5, DRPtheory2, DRPrate, Elber0} is based on the saddle-point approximation of the stochastic path integral (\ref{PI}) and 
focuses on such "biological instantons".
Unlike in QCD, such an approach is not affected by infrared divergences, hence it is completely rigorous. The expansion parameter controlling the accuracy of this  approximation is the thermal energy $k_B T$, which enters in the definition of the diffusion constant $D$ and of the effective potential $V_{eff}(x)$. 
 
In principle, the instantons can be obtained by relaxing numerically a discretized representation of the effective action functional $S_{eff}[x]$, in complete analogy with the cooling 
procedure developed in lattice field theory. In practice, however,  the simultaneous presence of fast and slow internal time scales (which range from fs to ns) makes such a task very challenging. 
 Indeed, the  discretized elementary time interval $\Delta t$ has to be chosen much smaller than the smallest time scale. Hence,  computing a single  dominant pathway would 
 require to find a minimum of a function 
with a gigantic number of degrees of freedom: $3 N \times N_t$, where $N$ is the number of atoms in the protein ---typically from several hundreds to many thousands--- and $N_t$ is the number 
of time discretization steps ---typically $10^6$---. 

Fortunately, a major numerical simplification of this problem can be achieved by exploiting the fact that the  dynamical system defined by the effective action (\ref{Seff}) is simplectic, i.e. 
it conserves the  "effective energy"
\footnote{Note that $E_{eff}$  has the dimension of a rate, therefore this quantity does not have the physical interpretation of a mechanical energy}
\be
E_{eff}= \frac{1}{4D} \dot{ x}^2(t) - V_{eff}[x(t)].
\ee
Hence, rather than minimizing directly the effective action $S_{eff}[x]$, it is possible to obtain the biological instantons
using the Hamilton-Jacobi (HJ) formulation of classical mechanics.  In other words, the trajectories obeying the classical equations of motion and subject to the 
boundary conditions $x(t)=x_f$, and $x(0)=x_i$
are those which minimize the effective HJ functional
\be
\label{SHJ}
S_{HJ}[x(l)]    = \frac{1}{\sqrt{D}}~\int_{x_i}^{x_f}d l \sqrt{E_{eff}(t) + V_{eff}[x(l)]},
\ee
where $dl = \sqrt{d x^2}$ is the measure of the distance covered by the system in configuration space,  during the transition. Hence, the dominant reaction pathways can be viewed as 
 the geodesic in a space
with a curved  metric.

The major advantage of the HJ formulation is that it allows to remove the time as the independent
variable and replacing it with the curvilinear abscissa $l$, which has the dimension of a length scale. Since there is no gap in the length scales of molecular systems, the discretization of the HJ is expected to converge extremely much faster than the time discretization of the
effective action $S_{eff}[x]$. Indeed, typically 100 slices are sufficient to achieve an accurate representation of the path. 

The effective energy $E_{eff}$ is an external parameter which determines the time at which each configuration of  a dominant reaction pathway $\bar x(\tau)$ is visited,  according to the usual HJ relationship
\be
\label{time}
t(x) =\int^{x}_{x_i} dl \frac{1}{ \sqrt{4 D(E_{eff}+V_{eff}[\bar x(l)])}}.
\ee
Typically, one is interested in studying transitions which terminate close the local minima of the potential energy $U(x)$. The residence time in such end-point configurations must be much longer than that in the configurations visited during the  transition. 
From Eq. (\ref{time}) it follows that these conditions are verified if
$
E_{eff}\sim -V_{eff}(x_o),
$
where $x_o$ is a configuration in the vicinity of a local minimum of $U(x)$.

In practice, computing the dominant reaction pathway connecting two given configurations $x_i$ to $x_f$ amounts to minimizing a discretized version of the effective HJ functional:
\begin{equation} 
\label{effact_discr}
 S_{HJ}^{d} [x(l)]= \sum_{n=1}^{N_s-1} \sqrt{ \frac{1}{D} \left[
E_{eff} + V_{eff}\left( x(n) \right) \right] } \; \Delta l_{n,
n+1},
\end{equation}
where $N_s$ is the number of path discretization slices. Once such a path has been determined, one can reconstruct the time at which each of the configurations is visited during the transition, using Eq. (\ref{time}). In particular, the time interval between the $n$-th and the $(n+1)$-th slice is 
\be
\label{times}
\Delta t_{n+1, n} = \frac{\Delta l_{n+1, n}}{ \sqrt{4 D(E_{eff}+V_{eff}[\bar x(n)])}},
\ee
where $\Delta l_{n+1, n}= \sqrt{(\bar x(n+1)-\bar x(n))^2}$. 
\subsection{Thermal Fluctuations Around the Dominant Folding Pathways}
\label{pra}

The biological instantons encode information about the reactive dynamics, hence can be used as starting point to compute the folding rate, whose inverse gives the typical 
time a protein takes to reach the native state, starting from a denatured state. This is a fundamental 
observable in protein kinetics, which is accessible from fluorescence experiments. 

It turns out that, in order to obtain an equation for the rate, it is necessary to go beyond the lowest-order saddle-point approximation, and estimate the  
contribution of small thermal  fluctuations around the dominant paths \cite{DRPrate}.

This corresponds to evaluating the instanton weight at one-loop level,  by functionally expanding the effective action to quadratic
order, around  the instanton solution $\ox(\tau)$:
\be
\label{DRPexpansion}
S_{eff}[x] &=& S_{eff}[\ox]+ \frac{1}{2} \int_{0}^{t} d\tau' \int_{0}^{t} d\tau'' 
\frac{\delta^2 S_{eff}[\ox]}{\delta x_i(\tau') \delta x_k(\tau'')}~(x_i(\tau')-\ox_i(\tau')) \,(x_k(\tau)-\ox_k(\tau))+ \ldots \nonumber\\
&\simeq& S_{eff}[\ox] + \frac{1}{2} \int_{0}^{t} d\tau~\int_{0}^{t} d\tau' (x_i(\tau')-\ox_i(\tau')) ~\hat{F}^{\tau, \tau'}_{i k}[\ox]~(x_k(\tau)-\ox_k(\tau)),
\ee
where we have introduced the fluctuation operator $\hat{F}[\ox]$, defined as 
\be
\hat{F}^{\tau, \tau'}_{i k}[\ox] \equiv \frac{\delta^2 S_{eff}[\ox]}{\delta x_i(\tau') \delta x_k(\tau'')} =
\left[-\frac{1}{2D}~\delta_{i k }~\frac{d^2}{d \tau^2} + \partial_i \partial_k~V_{eff}[\ox(t)]\right] \delta(\tau-\tau').
\ee

The formal one-loop expression for the conditional probability is therefore
\be
\label{KtWKB}
P(x_f,t|x_i)&\simeq&  \mathcal{N}~\frac{e^{-\frac{\beta}{2 }(U(x_f)-U(x_i))}}{\sqrt{\det \hat{F}[\ox]}}~e^{-S_{eff}[\ox]}.
\ee
Such an expression  is clearly divergent and needs to be regularized. To this end, one can multiply and divide by the conditional probability density  of a reference system $P_{reg.}(x_f, t|x_i)$. For example, one can use as regulator the conditional probability associated to the diffusion in an external harmonic potential, for which the propagator $P_{reg.}(x_0, t|x_0)$
 is analytically known. The result is \cite{DRPrate}:
\be
P_{DRP}(x_f,t|x_i)&=&  P_{reg.}(x_0, t|x_0)~e^{-\frac{\beta}{2 }(U(x_f)-U(x_i))}~e^{S^{reg.}_{eff}[\ox_{reg.}]-S_{eff}[\ox]}
~\sqrt{\frac{1}{\det \left(\hat{F}^{-1}_{reg.}[\ox_{reg.}]~\hat{F}[\ox]\right)}} \label{main},\nonumber\\
\ee
where $\hat F_{reg}$ is the fluctuation operator for the reference system used in the regularization. 

In practice, the  determinant $\det \left(  \hat{F}_{reg.}[\ox_{reg.}]~\hat{F}^{-1}[\ox]\right)$ has to be evaluated numerically from 
 the discretized representation of the fluctuation operators. To obtain such a representation,   one needs to express the time derivative using discretized time intervals.
 It is most convenient to use the intervals $\Delta t_{i, i+1}$ evaluated from the dominant trajectory, according to Eq. (\ref{times}). The fluctuation operator reads
\be
\label{discrF}
\hat{F}[\bar{x}]^{i,j}_{k,m} &=& \frac{-1/D}{\Delta t_{m+1,m}+\Delta t_{m,m-1}} \delta_{i,j} \, \left[\frac{\delta_{k,m+1}}{\Delta t_{m+1,m}}- \delta_{k,m} 
      \left( \frac{1}{\Delta t_{m+1,m}}+ \frac{1}{\Delta t_{m,m-1}} \right) +\frac{\delta_{k,m-1}}{\Delta t_{m,m-1}}  \right] \nonumber\\
 &+&   \frac{\partial^{2}\veff(\bar{x}(k))}{\partial x_{i} \partial x_{j}} \delta_{k,m},
    \ee
    where the indexes $k, m=1, \dots, N_s$ run over the path frames, while the indexes $i, j=1, \ldots, d$ label the degrees of freedom of the system. 

\subsection{Improved Effective Actions}

The main numerical advantage of the DRP formalism arises from the possibility of removing the time as a dynamical variable, and replace it with the curvilinear abscissa $l$. 
Since there is no gap in 
the characteristic length scales of molecular systems, the convergence of the discretization of $l$ is usually very fast compare to that of the discretization of $t$. 
As a result, using such a formulation, it
is possible to gain  information about the reaction mechanism at a very low computational cost. 
 
On the other hand, in order to obtain information about the dynamics, one needs to compute the times at which each configuration is visited along the dominant path, using Eq.s (\ref{time}) and (\ref{times}).  
The calculation of the time intervals $\Delta t_{i,i+1}$ is also needed in the discretized representation of the fluctuation operator (\ref{discrF}), which enters in the 
calculation of the order $k_B T$ corrections arising from non-equilibrium stochastic fluctuations around the dominant path. 
 
Clearly, the trick to remove time by switching  to the HJ formulation does not help in the evaluation of explicitly  time-dependent observables. 
As a result, in order to achieve an accurate description of the dynamics, or in order properly take into account  of the effects of fluctuations, one would  need to use a large number of path 
frames, with a consequent significant  increase of the computational cost of a DRP simulation. 

Fortunately,  the convergence of the calculation of the time intervals in the DRP approach can be greatly improved by adopting the effective stochastic theory (EST) 
developed in \cite{EST}.   The main idea of the EST is to exploit the gap in the internal time scales in order to analytically perform the integral over  the fast Fourier components of the  
paths $x(\tau)$ which contribute to the path integral (\ref{PI}), using Wilson's approach. Through such a procedure, the effects of the fast dynamics is rigorously and systematically integrated out
by renormalizing  the effective potential:
\be
\label{VR1}
V_{eff}(x) &\rightarrow& V^{EST}_{eff}(x) = V_{eff}(x) + V^{R}_{eff}(x).
\ee
where, to leading order in the effective theory one has
\be 
\label{VR2}
V^{R}_{eff}(x) &=& \frac{D \Delta t_c~(1-b)}{2 \pi^2 b} \nabla^2 V_{eff}(x) + \ldots
\ee 
In such an Eq., $\Delta t_c$ is a cut-off time scale which must be chosen much smaller than the fastest internal dynamical time scale and $b$ is a parameter which defines the interval of Fourier modes which are
 being analytically integrated out. Typically, for molecular systems  $\Delta t_c \sim 10^{-3}$ps and $b\sim 10^{-2}$ ~\cite{RGMD}.
 The dots in Eq. (\ref{VR2}) denote higher order correction in an expansion in the ratio of slow and fast time scales (so-called slow-mode perturbation theory) and can be found in the original publication \cite{EST}.

The EST generates by construction the same long-time dynamics of the original "bare" theory, but has a lower time resolution. In the context of MD simulations, this implies that the EST can be integrated using much larger discretization time steps~\cite{RGMD}. 
In the  context of the  DRP simulations, the utility of the EST resides in the fact that much fewer path discretization time steps are required in order to achieve a convergent calculation 
of the time interval from the dominant path, through Eq. (\ref{time}). 

\subsection{Generalization to Ab-Initio Quantum Chemistry Calculations} 
\label{QDRPsect}
So far we have discussed the DRP formalism in the specific context of protein folding. On  the other hand, this theory applies in general to any physical system described by the over-damped Langevin Eq. (\ref{Lan2}). In particular, it can be applied to investigate rare chemical 
reactions taking place in solution~\cite{QDRP1}.  Typically, such reactions involve large modifications of the electronic structure, e.g. associated to the breaking or forming of 
covalent bonds. Hence, in investigating such processes, a classical description based on some phenomenological expression for the molecular potential energy $U(x)$ entering
in the Langevin equation is no longer appropriate, and one must adopt a sub-atomic approach, in which the quantum electronic degrees of freedom are explicitly taken into account.
 
In the Born-Oppenheimer approximation~\cite{Leach}, this is done in two steps. First, the Schr\"dinger equation for  the electronic degrees of freedom is (approximatively) 
solved holding fixed all the nuclear coordinates. The corresponding ground-state energy is then interpreted as the potential molecular energy $U(x)$ and is used to derive the 
effective potential $V_{eff}(x)$, which enters the DRP formulas. This approach allows to rigorously take into account of the quantum effects on the electron dynamics, to all order in  
$\hbar$ and to leading
order in the ratio $m/M$ between the mass of the electron $m$ and  the typical mass of the atomic nuclei $M$.  

\subsection{Quantum Corrections to the Diffusive Dynamics of the Atomic Nuclei}

In the discussion performed so far,  the dynamics of the atomic nuclei has been assumed to be entirely classical, and to evolve according to the Langevin Eq. (\ref{Lan2}). 
Such an approximation is certainly reliable for most atomic species which are found in biomolecular systems, such as carbon or oxygen. On the other hand, it becomes questionable
for the lightest species, notably hydrogen. Hence, in the study of reactions involving a large number of such atoms, the quantum corrections to the stochastic dynamics of nuclei  should be consistently taken into account. 

To tackle this problem, we have derived the multidimensional generalization~\cite{QLDRP} of the  semiclassical extension of the Langevin Eq. (\ref{Lan2}) and of the corresponding FP equation (\ref{FPE}), 
in which quantum corrections are  systematically included to order $\hbar^2$ \cite{QSE, Coffey}. From these equations it is immediate to compute 
the quantum corrections to the DRP equations~\cite{QLDRP}. 
We found that the  effective action receives contribution from an additional term in the form (adopting again for simplicity a one-dimensional notation):
\be
S_{eff}^Q[x] = \int_0^t d \tau ~V_{eff}^Q[x(\tau)],
\ee
where
\be
\label{VQeff}
V_{eff}^Q(x) \simeq  \frac{D~\lambda}{4 (k_B T)^3} |\nabla U(x)|^2 \nabla^2 U(x)
\ee 
is the correction to the effective potential and
$\lambda= \frac{\hbar^2}{12 ~k_B T m_i}$ is a characteristic parameter which determines the size of quantum fluctuations. In conformational transitions driven by the formation 
of hydrogen bonds, the quantum corrections to the dominant reaction pathways have been found to give raise to sizable effects, see Ref.~\cite{QLDRP}.
 
\section{Some Applications}

The DRP approach has been extensively tested and validated on toy systems~\cite{DRPtheory2, DRPrate} and simplified protein models~\cite{DRPtest1, DRPtest2}, and then applied to investigate 
realistic transitions, including conformational transitions of peptides~\cite{DRPtheory1.5, QDRP2}, chemical reactions~\cite{QDRP1} and realistic protein folding reactions~\cite{FoldingFIP}. 
In this section, we briefly review some of these studies. 

\subsection{Validation of the DRP Theory on a Coarse-Grained Model}

The most straightforward way to assess the accuracy of the DRP approach consists in  comparing its predictions with those obtained directly 
from MD simulations. Unfortunately, for realistic models MD simulations of protein folding reactions are not presently feasible.  

\begin{figure}[t]
 \center{
  \includegraphics[width=9 cm]{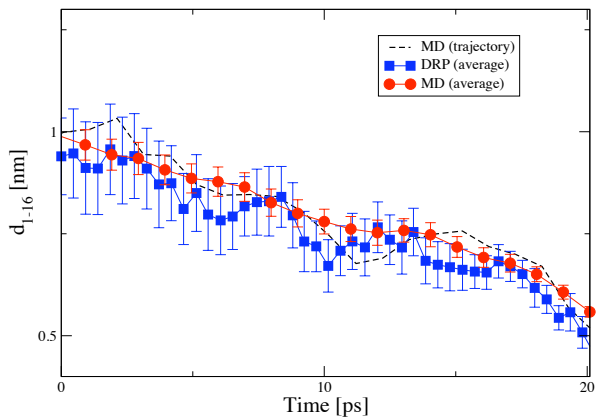}
  \caption{Evolution of the end-point distance of the protein during a folding reaction, averaged over folding trajectories obtained in DRP and MD simulations, respectively. The dashed line
  represents a typical trajectory.}\label{compare1}
}
    \end{figure}

Hence, we have validated the DRP approach using much simpler coarse-grained models, with protein-like properties. For example, the model used 
in Ref.\cite{DRPtest2} displays a folding transition with a well-defined and unique native state. 
The predictions for the time evolution of the distance between the end-points of the chain (a typical observable in protein folding)  obtained in the DRP and MD approaches is compared in Fig. \ref{compare1}. We see that the biological instanton standpoint provides a realistic description of the reactive dynamics. 
 
\subsection{Atomistically Detailed Protein Folding Simulations}
The computational efficiency of the DRP approach allows to investigate a large class of reactions which cannot be studied on the existing supercomputer using standard MD simulations. 
Only very few examples of successful MD simulations of protein folding reactions have been reported to date. These studies were either performed on a special purpose 
supercomputing machine in which 
all the key MD algorithms are implemented at the hardware level~\cite{Anton} or through very long large-scale simulations performed on several thousands of CPU's,  
distributed world wide~\cite{Pande}. 

On the other hand, using the DRP method~\cite{FoldingFIP} it was possible to simulate the same reaction at a comparable level of statistics in less than a single day on only 32 CPU's. The DRP 
results were found to be consistent with those reported in \cite{Anton}, and provided an explanation of the discrepancy between the results of Ref.~\cite{Anton} 
and Ref.~\cite{Pande}. An illustration of one of the folding trajectories obtained in the DRP approach is displayed in Fig. \ref{FoldingFIP}. Simulations on much larger systems
(unaccessible to MD simulations) are now being performed. 

\subsection{Folding of a Small Peptide from Ab-Initio Quantum Mechanical Simulations} 
\begin{figure}[t]
\center{
  \includegraphics[width=10 cm]{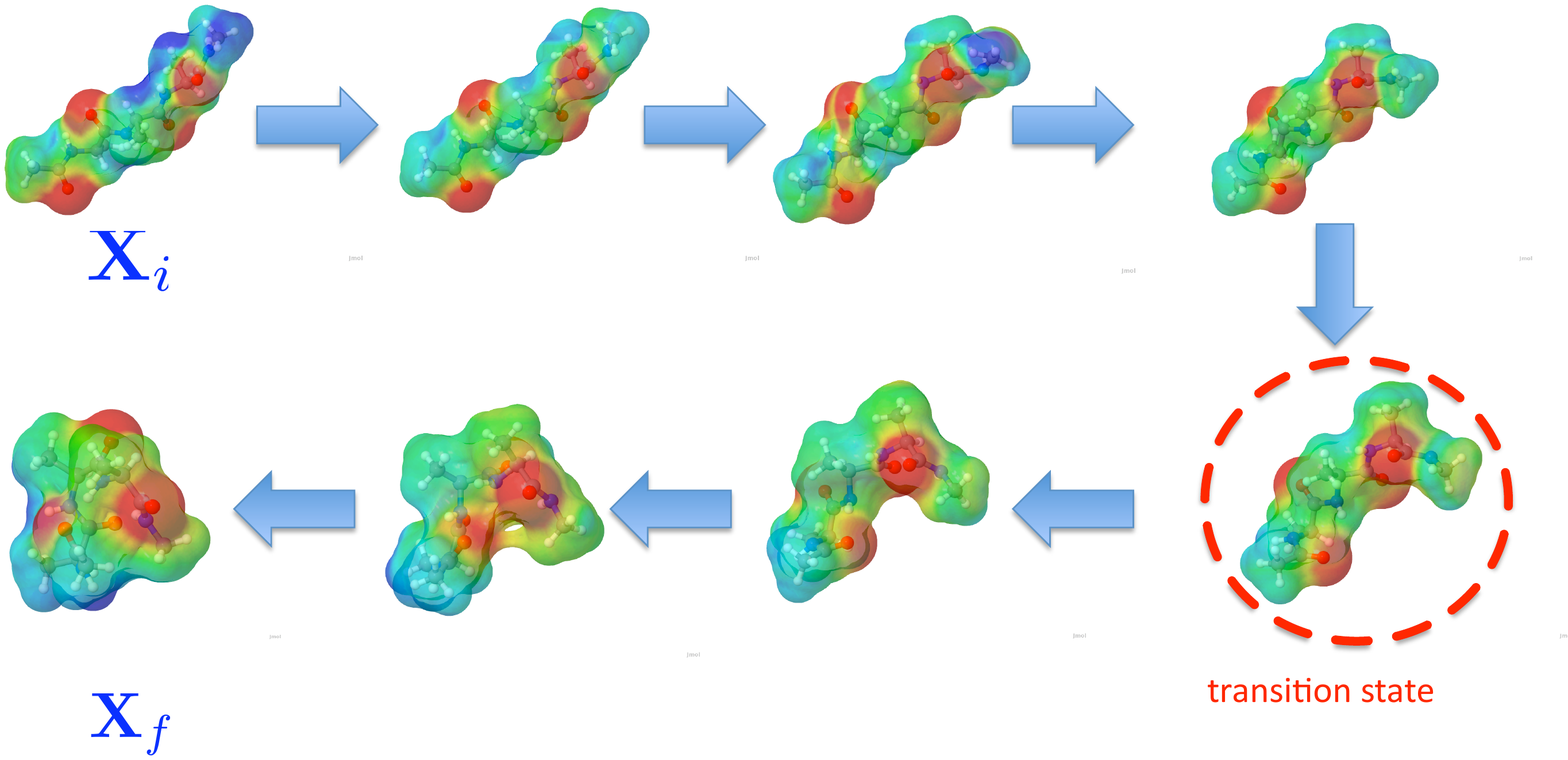}
  \caption{Folding of a peptide obtained from quantum DRP calculations. The shaded area represents the distribution of electric charge.} \label{QDRPfig}
}
   \end{figure}

Using the DRP approach we have performed the first study of the folding of a peptide based on \emph{ab-initio} Quantum Mechanical calculations, using the Born-Oppenheimer 
 approach described in section \ref{QDRPsect}. This type of analysis could never performed using MD or Car-Parinello approaches~\cite{Leach}. 
The results of our simulations are schematically represented in Fig.~\ref{QDRPfig}. This study allowed for the first time to access the reliability of the existing classical force fields 
in predicting non-equilibrium reactive trajectories. This is an important and non-trivial issue, since the phenomenological force fields are fitted in order to reproduce quantum mechanical calculations
on static configurations, or in equilibrium condition. Hence, there is in principle no guarantee that they should be accurate in also in the regions of phase-space which are mostly 
visited by the non-equilibrium reactive trajectories. 
On the other hand, our studies have revealed that classical  calculations give results that are indeed in good agreement with those of quantum \emph{ab-initio} simulations. 
This study also allowed to quantify to what extent the electronic structure of the chain is modified during the reaction. 
 
\section{Conclusions}

Nuclear physics  provides an excellent training ground where developing advanced theoretical approaches to   interacting many-particle 
systems.  
Indeed, a number of schemes and approximations 
originally proposed in this context~\cite{negeleorland} have later become standard tools in the field of strongly correlated electrons, quantum liquids, cold atoms, \ldots.

In particular, in this talk we have reviewed our recent attempt to export functional integral techniques originally derived in QCD to investigate 
the non-equilibrium diffusive dynamics of biomolecules such as proteins. 
We have argued that such an approach opens the door to
the study of a large number of processes which simply could not be investigated using standard MD simulations. 

Interdisciplinary studies of this type may also emphasize the importance of pushing forward the fundamental research in nuclear theory. Indeed, they show
that this activity is valuable also in a perspective which goes beyond the  traditional perimeter of nuclear physics.  

\section*{Acknowledgments}
I would like to thank all the collaborators which have been involved in the development of the DRP approach, and in particular H. Orland, F. Pederiva, G. Garberoglio, S. a Beccara and M. Sega. 

PF is a member of the Interdisciplinary Laboratory for Computational Science (LISC), a joint venture of Trento University and Bruno Kessler foundation. The numerical simulations were performed 
on the Aurora supercomputer at LISC. This research was partially funded by the Provincia Autonoma di Trento and by INFN, through the AuroraScience project.


\section*{References}
\begin{thebibliography}{99}
\bibitem{Freddolino} P. L. Freddolino \emph{et al.}, Biophys. J. {\bf 94}, L75 (2008).
\bibitem{proteinphysics} A.V. Finkelstein and O.B. Ptitsyn, "Protein Physics: a Course of Lectures", Academic Press, London (2002).
\bibitem{Leach} A.R. Leach,  "Molecular modeling: principle and applications" (2nd ed.) Pearson Education (Harlow, England), 2001. 
\bibitem{Schwabl} F. Schwabl, "Statistical Mechanics" (2nd ed.), Springer-Verlag, Berlin 2006. 
 \bibitem{DRPtheory1} P. Faccioli, M. Sega, F. Pederiva  and H. Orland, Phys. Rev. Lett. {\bf 97}, 108101 (2006).   
 \bibitem{DRPtheory1.5} M. Sega, P. Faccioli, F. Pederiva, G. Garberoglio and H. Orland, Phys. Rev. Lett. {\bf 99}, 118102 (2007).
 \bibitem{DRPtheory2} E. Autieri, P. Faccioli, M. Sega, F. Pederiva  and H. Orland,  J. Chem Phys. {\bf 130}, 064106 (2009).  
\bibitem{DRPrate} G. Mazzola, S. a Beccara, P.Faccioli, and H. Orland,  J. Chem. Phys. {\bf 134}, 164109 (2011).
\bibitem{Elber0} R. Elber,  and D. Shalloway, J. Chem. Phys. {\bf 112}, 5539 (2000).
\bibitem{EST} O.~Corradini, P.~Faccioli and H.~ Orland, Phys. Rev. {\bf E80} 061112 (2009) .
\bibitem{RGMD} P. Faccioli,  J. Chem. Phys. {\bf 133} 164106 (2010).
\bibitem{QDRP1} S. a Beccara, G. Garberoglio, P. Faccioli and F. Pederiva, J. Chem. Phys. {\bf 132}, 111102 (2010).
\bibitem{QSE} J. Ankerhold, P. Pechukas and H. Grabert, Phys. Rev. Lett. {\bf 87}, 086802 (2001). J. Ankerhold and H. Grabert, Phys. Rev. Lett. {\bf 101}, 119903 (2008) (Erratum). 
J. Ankerhold, Phys. Rev. {\bf E64}, 060102 (2001). 
\bibitem{Coffey}  W. T. Coffey, Y. P. Kalmykov, S. V. Titov, and B. P. Mulligan, J. Phys. {\bf A 40}, F91(2007).  W. T. Coffey, Y. P. Kalmykov, S. V. Titov, AND L. Cleary, Phys. Rev. {\bf E 78} 031114 (2008).   
\bibitem{QLDRP} S. a Beccara, G. Garberoglio and P. Faccioli, J. Chem. Phys. {\bf 135} 034103 (2011).
\bibitem{DRPtest1} P. Faccioli, J. Phys. Chem. {\bf B112}, 137560 (2008).
\bibitem{DRPtest2}  P. Faccioli, A. Lonardi and H. Orland,  J. Chem. Phys. {\bf 133}, 045104 (2010).
\bibitem{QDRP2} S. a Beccara, P. Faccioli, M. Sega, G. Garberoglio, F. Pederiva and H. Orland, J. Chem. Phys. {\bf 134}, 024501 (2011).
\bibitem{Anton} D.E. Shaw \emph{et al.}, Science {\bf 330}, 341 (2010).
\bibitem{Pande} D.E. Ensign and V.J. Pande, Biophys. Journ. {\bf 96} L53 (2009).
\bibitem{FoldingFIP} S. a Beccara, T. Skrbic, R. Covino and P. Faccioli, presently under review. 
\bibitem{negeleorland} J.W. Negele and H. Orland, "Quantum many-particle systems",  Westview Press 1998.
\end{thebibliography}
\end{document}